\pdfoutput=1

\documentclass[11pt]{article}

\usepackage[final]{acl}

\usepackage{times}
\usepackage{latexsym}

\usepackage[T1]{fontenc}

\usepackage[utf8]{inputenc}

\usepackage{microtype}

\usepackage{inconsolata}

\usepackage{times}
\usepackage{soul}
\usepackage{url}
\usepackage{graphicx}
\usepackage{amsmath}
\usepackage{amsthm}
\usepackage{booktabs}
\usepackage{algorithm}
\usepackage{algorithmic}
\usepackage{
subfigure}
\usepackage{makecell}
\usepackage{multirow}
\usepackage{color}
\usepackage{enumerate}

\title{StreamVoice: Streamable Context-Aware Language Modeling for Real-time Zero-Shot Voice Conversion}

\author{
Zhichao Wang$^1$,
Yuanzhe Chen$^2$,
Xinsheng Wang$^1$,
Lei Xie$^1$\footnotemark[1],
Yuping Wang$^2$
\\    
$^1$Audio, Speech and Language Processing Group (ASLP@NPU)\\School of Computer Science, Northwestern Polytechnical University, Xi’an, China\\
$^2$Douyin Vision Co., Ltd.\\
}

\begin{document}
\maketitle

\renewcommand{\thefootnote}{\fnsymbol{footnote}}
\footnotetext[1]{Corresponding author}
\renewcommand{\thefootnote}{\arabic{footnote}}
\begin{abstract}
Recent language model (LM) advancements have showcased impressive zero-shot voice conversion (VC) performance. However, existing LM-based VC models usually apply offline conversion from source semantics to acoustic features, demanding the complete source speech and limiting their deployment to real-time applications. In this paper, we introduce StreamVoice, a novel streaming LM-based model for zero-shot VC, facilitating real-time conversion given arbitrary speaker prompts and source speech. 
Specifically, to enable streaming capability, StreamVoice employs a fully causal context-aware LM with a temporal-independent acoustic predictor, while alternately processing semantic and acoustic features at each time step of autoregression which eliminates the dependence on complete source speech. 
To address the potential performance degradation from the incomplete context in streaming processing, we enhance the context-awareness of the LM through two strategies: 1) teacher-guided context foresight, using a teacher model to summarize the present and future semantic context during training to guide the model's forecasting for missing context; 2) semantic masking strategy, promoting acoustic prediction from preceding corrupted semantic and acoustic input, enhancing context-learning ability. Notably, StreamVoice is the first LM-based streaming zero-shot VC model without any future look-ahead. Experiments demonstrate StreamVoice's streaming conversion capability while achieving zero-shot performance comparable to non-streaming VC systems.
\end{abstract}

\begin{figure}[h]	
	\centering
    	\includegraphics[width=0.9\linewidth]{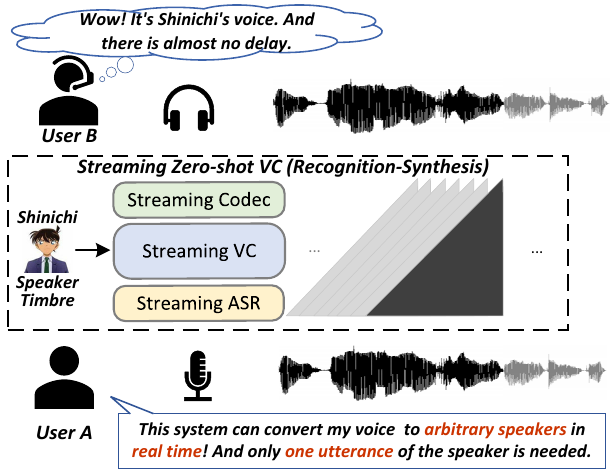}
 \caption{The concept of the streaming zero-shot VC employing the widely used recognition-synthesis framework~\cite{PPGSun2016PhoneticPF}, where only the encoder of ASR is involved. StreamVoice is built on this popular paradigm.}
	\label{fig:recong-systh}
 \vspace{-15pt}
\end{figure}

\section{Introduction}
\label{sec:intro}

Voice conversion (VC) aims to transfer a speaker's voice to that of another speaker without changing the linguistic content. This technique has been deployed in many real-world applications, such as movie dubbing, privacy protection, pronunciation correction, etc. 
With the help of neural semantic features, such as the bottleneck feature (BNF) from an automatic speech recognition (ASR) system, converting source speech from arbitrary speakers in the wild has been successfully achieved~\cite{PPGSun2016PhoneticPF}. 
Meanwhile, converting to an arbitrary target speaker with only one utterance of this speaker, called \textit{zero-shot VC}, has also been researched recently~\cite{autovcqian2019autovc,LM-VC}. 
However, most existing zero-shot VC models are designed for offline systems, which are insufficient to meet the recent growing demands of streaming capability in real-time VC applications, such as live broadcasting and real-time communication (RTC). In this study, we focus on the \textit{streaming zero-shot VC} as illuminated in Fig.~\ref{fig:recong-systh}.

Disentangling speech into different components, e.g., semantic content and speaker timbre, plays an important role in the zero-shot VC task~\cite{INchou2019oneshot,wang2023multilevel,VQMIVC,autovcqian2019autovc}. Recently, 
benefiting from the powerful LM framework and the scaling up of training data, LM-based VC models~\cite{LM-VC,uniaudio,vectok} with built-in in-context learning ability can learn the context relations between source and target speaker's utterances to
capture fine-grained speaker timbre, achieving impressive zero-shot VC performance.
However, demanding the complete source speech utterance limits these LM-based VC models to real-time scenarios; thus, they can only be used in offline applications.
While several non-LM-based methods~\cite{liveosvc,ALO-VC} have been proposed for streaming zero-shot VC, the performance fails to generalize well to unseen speakers with high speaker similarity and speech naturalness, mainly due to the limited model capacity to scale up training data, and also the performance degradation caused by the missing future information in streaming scenario.

Inspired by the success of LM-based models in zero-shot VC, we aim to explore the feasibility of LMs for the streaming VC scenario. An intuitive way is to follow the popular recognize-synthesis framework shown in Fig.~\ref{fig:recong-systh}, in which speech is represented in semantic BNF and acoustic features respectively extracted by a streaming ASR and an audio codec.
Then, the LM-based VC model undertakes the transformation of semantic information into acoustic features with the target speaker's timbre. However, the development of the LM-based model in streaming zero-shot VC is hampered by two primary challenges.
\begin{enumerate}[0]
    \item[$\bullet$]
    \textbf{Streamable architecture:} streaming models typically produce immediate output upon receiving current input without reliance on future time steps. Current LM-based VC models perform the conversion only when they get a full-utterance of source speech, which fails to meet the demands of streaming applications. The widely adopted multi-stage language modeling for multi-layer codec prediction introduces complexity to system design, posing a potential risk of cumulative errors. Additionally, the dependency models of the streaming pipeline also impact the design and performance of the VC model.
    \item[$\bullet$]
    \textbf{Performance gap:} unlike non-streaming models, streaming models must process frame-wise or chunked input causally on the fly without future information, facing missing context and potential performance degradation. This missing hinders the streaming VC model from achieving high-quality conversion. In addition, as shown in Fig.~\ref{fig:recong-systh}, the VC model relies on the semantic feature BNF from ASR to achieve conversion, which makes semantic features very important. However, streaming ASR exhibits inferior performance compared to its non-streaming counterpart, leading to the BNF carrying low-quality semantic information but more speaker information. In addition to the inherent unavailable future reception, this low-quality semantic input makes achieving high-quality conversion more difficult. The goal of zero-shot VC amplifies the challenges faced by our streaming VC model.
\end{enumerate}

In this work, we propose \textit{StreamVoice}, a streaming LM-based model for high-quality zero-shot VC.
Specifically, StreamVoice has a streamable architecture that integrates a single-stage language model that casually generates acoustic codecs with the collaboration of an acoustic predictor. Alternating input of semantic and acoustic features at each time step ensures seamless streaming behavior. Two methods are introduced to enhance the context-awareness of the LM to mitigate the performance gap caused by missing contextual information. 1) We incorporate a teacher-guided context foresight, where the VC model is taught by a teacher non-streaming ASR to infer the present and future semantic information summarized by the teacher, which is then used to enhance the acoustic prediction. 2) To enhance the context learning from the input history, semantic masking encourages acoustic prediction from the preceding acoustic and corrupted semantic input, which also implicitly creates an information bottleneck to reduce the source speaker's information.

Experiments demonstrate StreamVoice’s ability to convert speech in a streaming manner with high speaker similarity for both seen and unseen speakers while maintaining performance comparable to non-streaming VC systems. As the first LM-based zero-shot VC model without any future look-ahead, the total pipeline only has 124 ms latency to perform the conversion, 2.4x faster than real-time on a single A100 GPU without engineering optimizations. Converted samples can be found in \href{https://kerwinchao.github.io/StreamVoice/}{\url{https://kerwinchao.github.io/StreamVoice/}}.

\section{Related Works}



\noindent\textbf{Zero-shot Voice Conversion.}
Zero-shot VC imposes stringent demands on speech decoupling and capturing speaker timbre. Many studies specifically design many disentanglement approaches, incorporating intricate structures~\cite{INchou2019oneshot}, loss functions~\cite{VQMIVC}, and training strategies~\cite{contrastive}, to achieve speech decoupling. Rather than embedding explicit disentanglement designs in VC training, some approaches~\cite{mediumvc} leverage a speaker verification (SV) model for speaker representation, while linguistic content is extracted using ASR or self-supervised learning (SSL) models~\cite{PPGSun2016PhoneticPF,nansy}. To enhance speaker timbre capturing, some fine-grained speaker modeling methods have also been explored~\cite{retriever,wang2023multilevel}. Recent successes of language models in generative tasks have prompted the exploration of LM-based models in zero-shot VC, yielding impressive results. Using the pre-trained model to decouple speech, the LM-based VC model~\cite{LM-VC,uniaudio,vectok} captures fine-grained speaker timbre from the speaker prompt and then performs the conversion. However, current LM-based VC models are inapplicable to streaming scenarios, constraining their real-world utility. This paper addresses this gap by investigating the zero-shot capabilities of language models specifically tailored for streaming scenarios.

\noindent\textbf{Streaming Voice Conversion.}
Despite the high-quality conversion achieved by non-streaming VC models, their non-streamable structure and reliance on full-utterance input hamper them for real-time streaming applications. For streaming, causal processing and the structure of the streaming pipeline are crucial considerations. Streaming models are compelled to process frame-wise or chunked input on the fly, devoid of access to future information, leading to performance degradation compared to non-streaming counterparts. To address this, a common approach~\cite{investgateStreaming,kameoka2021fasts2s,dualvc,dualvc2} involves the integration of a teacher model to guide the training of the streaming model or the distillation of knowledge from a non-streaming model. Chen et al.~\shortcite{YZ} focus on selecting BNF with minimal semantic information loss through layer-wise analysis, while Chen et al.~\shortcite{chen2022streaming} incorporate adversarial training to enhance the quality of semantic features. Beyond streaming VC, some efforts have recently been towards streaming zero-shot VC. For instance, VQMIVC~\cite{VQMIVC}, designed for the non-streaming application, is modified to be streamable by Yang et al.~\shortcite{liveosvc}. ALO-VC~\shortcite{ALO-VC} constructs a streaming system using an SV model, a streaming PPG extractor, and a pitch extractor. However, current streaming zero-shot VC, designed for low-resource devices, has limited model capacity with poor generalization to unseen speakers, leading to inferior similarity and naturalness. Motivated by LM's successes in zero-shot VC, we design a streamable LM in streaming scenarios. To tackle distinctive challenges in streaming VC, we 
enhance the context awareness of the LM to improve conversion quality.


\noindent\textbf{Language Model-based Speech Generation.}
Recent advancements in LMs within natural language processing have showcased potent generation capabilities, influencing the development of LMs in speech generation. By employing codec~\cite{soundstream} or other SSL models~\cite{w2vbert}, speech and audio can be efficiently tokenized into discrete units, facilitating low-bitrate audio representation and semantic extraction. This progress allows speech generation to utilize LM frameworks seamlessly. Taking audio generation as a conditional language modeling task, AudioLM~\shortcite{audiolm} employs hierarchical language modeling for acoustic prediction from coarse to fine units. VALL-E~\shortcite{valle} and SpearTTS~\shortcite{speartts} extend LMs for zero shot-TTS, which can clone a human's voice with prompt tokens from a short recording. For zero-shot VC, LM-VC~\shortcite{LM-VC} employs task-oriented optimizations to this task. And some studies~\cite{vectok,uniaudio} leverage multitask objectives and datasets, achieving high-quality conversion. Despite this progress, existing LM-based VC models usually apply offline processing, demanding complete utterance from the source speech, which hinders their suitability for real-time streaming applications. In contrast to prior studies, we explore the zero-shot capability of the LM-based VC for streaming scenarios. With the enhancement of context awareness, the proposed LM-based VC model achieves results comparable to non-streaming LM-based VC.

\begin{figure}[htb]	
 \centering
    \includegraphics[width=0.9\linewidth]{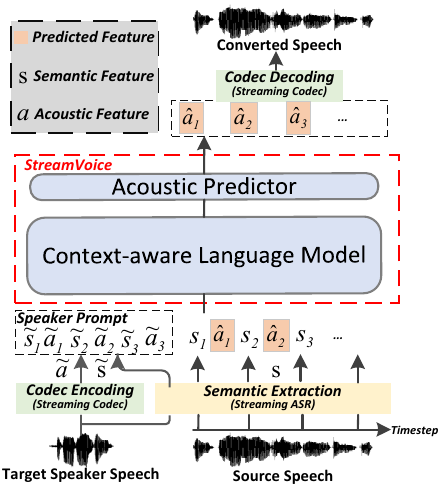}
 \caption{The overall architecture for StreamVoice.}
	\label{fig:streamvoice}
 \vspace{-10pt}
\end{figure}


\section{StreamVoice}

\subsection{Overview}
    As shown in Fig.~\ref{fig:streamvoice}, the development of StreamVoice follows the recognition-synthesis framework. In this framework, speech is first represented as semantic features $\mathbf{s} =\{s_1,s_2,...s_{T_s}\}$ and acoustic features $\mathbf{a} =\{a_1,a_2,...,a_{T_a}\}$ by a pre-trained streaming ASR model and a speech codec model respectively. Here, $T_s$ and $T_a$ denote the sequence length. Before inputting to StreamVoice, $\mathbf{s}$ and $\mathbf{a}$ are aligned to the same length $T$. StreamVoice incorporates a context-aware language model and an acoustic predictor to perform a single language modeling process. With the semantic and acoustic features $\{\mathbf{\tilde{s}},\mathbf{\tilde{a}}\}$ of speech from the target speaker as speaker prompt, the LM leverages the semantic information $\mathbf{s}_{1:t}$ of source speech to autoregressively predict the hidden output $^c\mathbf{h}$. In each autoregression time-step of the LM, the acoustic predictor transforms the hidden output $^c\mathbf{h}$ to the codec feature $\mathbf{\hat{a}}$ of the converted speech. Finally, the codec model reconstructs the waveform from the predicted codec feature. In the following sections, we will introduce how to build a streamable LM for VC and how to ensure the high-quality conversation of this streaming VC.

\begin{figure}[htb]	
	\centering
    	\includegraphics[width=1\linewidth]{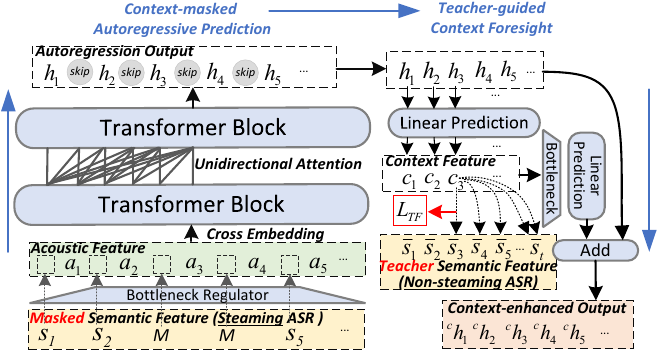}
 \caption{The architecture for context-aware LM.}
	\label{fig:lm}
\end{figure}



%

\subsection{Streamable Architecture} 

To perform streaming VC, a streamable architecture is necessary. In StreamVoice, the language model is carefully designed to perform full causal processing in the VC task, and the acoustic predictor is designed to achieve frame-wise prediction without dependency on temporal information.

\subsubsection{Fully Casual Language Model}
As shown in Fig.~\ref{fig:lm}, inspired by the success of the LM-based VC model, we intend to achieve streaming zero-shot VC by language models. In previous LM-based VC models~\cite{LM-VC}, the demand of the complete semantic feature $\mathbf{s}$ from source speech to achieve conversion hinders the deployment for real-time application, which can be formulated as $p(a_t|\mathbf{s}_{1:T_s},\mathbf{a}_{1:t-1})$ for each time step. To achieve streaming, any components of the LM cannot rely on future information. As shown in  Fig.~\ref{fig:lm}, decoder-only LM with unidirectional attention can easily fit the requirement of casual generation. To eliminate the dependency of the complete semantic input, semantic and acoustic features $\{\mathbf{s},\mathbf{a}\}$ are first aligned with each other to the same sequence length $T$ and then they are alternatively inputted to the LM, forming a cross-embedding like $\{s_1,a_1,s_2,a_2,...,s_T,a_T\}$. With these modifications, the LM can achieve streaming processing, modeling $p(a_t|\mathbf{s}_{1:t},\mathbf{a}_{1:t-1})$. 


To be specific, the semantic feature $\mathbf{s}$ obtained via an ASR model comprises a sequence of embeddings, denoted as $\{s_1, s_2, ..., s_T\}$. On the other hand, the codec tokens obtained from an $L$-layer codec are discrete units represented by $\mathbf{a} \in \mathcal{R}^{T \times L}$. To obtain the acoustic embedding sequence, the codec tokens from each layer undergo separate embedding into the embedding space, and then they are concatenated along the embedding dimension, resulting in the fused acoustic embedding. Both the fused acoustic embedding and semantic features are transformed into the same dimension using linear layers. Subsequently, they are alternately inputted into the language model, forming the cross-embedding.

\begin{figure}[htb]	
	\centering
    	\includegraphics[width=1\linewidth]{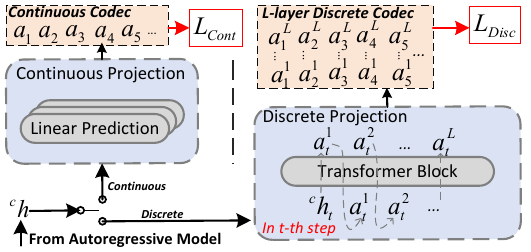}
 \caption{The architecture for acoustic predictor. Our system can support continuous or discrete projection.}
	\label{fig:codecpre}
\end{figure}

\subsubsection{Acoustic Predictor}


As the preceding LM has essentially encoded content and speaker into its output $^c\mathbf{h}$, the acoustic predictor can be designed in temporal irrelevant to transform $^c\mathbf{h}$ into acoustic codec space, which means the predictor can be easily applied in the streaming scenario. Given that the speech can be represented in acoustic features by neural codec in either continuous or discrete forms, we investigate the incorporation of both features in StreamVoice, which are performed by continuous projection and discrete projection, respectively. 


\textbf{Continuous Projection.}
Following Shen et al.~\shortcite{ns2}, the $D$-dimensional quantized latent vector $\mathbf{a} \in \mathcal{R}^{T \times D} $ encoded by the codec model is used as the continuous acoustic representation. The prediction of the continuous representation involves employing a stack of linear layers, as shown in Fig.~\ref{fig:codecpre}. The continuous projection loss is calculated as the L2 distance between the predicted acoustic feature $\hat{\mathbf{a}}$ and the ground-truth acoustic feature $\mathbf{a}$, which is defined as:
\begin{equation}
    \mathcal{L}_{Cont} = ||\mathbf{a}-\hat{\mathbf{a}}||^{2}_{2}.
\end{equation}

\textbf{Discrete Projection.}
In general, the codec is designed with multi-layer quantizers to compress original speech into $L$-layer discrete indices $\mathbf{a} \in \mathcal{R}^{T \times L}$ at a low bitrate. Most LM-based work~\cite{valle,LM-VC} stacks multiple LMs to predict discrete features, making the pipeline complicated and unsuitable for the streaming scenario. 
In contrast, StreamVoice adopts a streamlined multi-layer codec prediction method inspired by MQTTS~\cite{MQTTS}. This method, free from temporal dependencies, can seamlessly integrate into the streaming process of the language model.
Specifically, a single-layer transformer is used to model the hierarchical conditional distribution of codecs. As depicted in the right of Fig.~\ref{fig:codecpre}, at time $t$, the transformer employs the $^c\mathbf{h}$ as the starting condition and sequentially generates $a^l_t$ from layer 1 to L. Remarkably, this generation process is independent of the preceding or the future $\mathbf{^c\mathbf{h}}$, rendering it well-suited for the demands of a streaming scenario. Notably, in the proposed StreamVoice, we mainly incorporate the discrete projection to achieve acoustic prediction. The discrete projection loss can be described as:
\begin{equation}
\small
    \mathcal{L}_{Disc} = -log{\prod_{t=1}^{T} \prod_{l=1}^{L}p(a^l_t|\mathbf{a}_{1:t-1},^m\mathbf{s}_{1:t},t,a^{1:l-1}_{t})}. 
\end{equation}

\subsection{Context-aware Enhancement}

Due to the disadvantage of the causality in the streaming framework, streaming models face missing future reception and potential performance degradation compared to the non-streaming model, while the low-quality semantic input from the streaming ASR, as we mentioned in Section~\ref{sec:intro}, makes achieving high-quality conversion more challenging. 
To address these issues, a context-aware enhancement method is proposed, which can alleviate incomplete contextual information arising from the semantic input and the absence of future information.
Specifically, we introduce context-masked autoregressive prediction in the LM to enhance the capture of \textit{historical} context from the given semantic input. Meanwhile, a teacher-guided context foresight is proposed to ensure the model can imagine the \textit{future} context based on that of its historical context.


\textbf{Context-masked Autoregressive Prediction.}
As shown in the left of Fig.~\ref{fig:lm}, the LM is achieved by the multi-layer Transformer with unidirectional attention, following the implementation of LLaMA~\cite{llama}. To enhance contextual awareness from the given semantic input, semantic masking is introduced in the LM to encourage acoustic prediction from the corrupted semantic. Specifically, within a sequence of semantic tokens $\mathbf{s}=\{s_1,s_2,...s_{T}\}$, we randomly select several indices as start indices at a ratio $r$, and spans of $l$ steps are masked by \textit{[M]}. After masking, LM takes the corrupted semantic feature $^{m}\mathbf{s}$ as input and performs autoregression. With this method, an information bottleneck is also implicitly created in the semantic feature to reduce speaker information. Moreover, during training, we do not explicitly use a speech clip as the speaker prompt. Instead, LM leverages the previous sequence $\{\mathbf{s}_{1:t-1},\mathbf{a}_{1:t-1},s_t \}$ as prompts to autoregressively generate hidden representation $h_{t}$ for further acoustic prediction. Notably, when the current input is $a_t$, the corresponding output is skipped and does not involve further steps.

\textbf{Teacher-guided Context Foresight.}
As previously discussed, the absence of future information resulting in the loss of contextual information leads to a decline in the conversion performance. Inspired by the effective representation learning exhibited by autoregressive predictive coding~\shortcite{apc} (APC), we introduce teacher-guided context foresight guided by a non-streaming ASR to enhance the autoregression output, as presented in the right of Fig.~\ref{fig:lm}. This allows the model to learn a context vector containing envisioned future information. Specifically, the context representation $\mathbf{c}$ is first derived by linear prediction from the hidden features $\mathbf{h}$, which is generated by the LM through historical context. Subsequently, this $c_t$ is encouraged to discover more general future context information by minimizing the L2 distance not only with $k$ semantic features from future time steps ${\overline{s}_{t+1},...,\overline{s}_{t+k}}$ but also with the current semantic $\overline{s_t}$. This dual minimization approach contributes to precise content delivery and enhances the ability to forecast future context. The loss can be summarized as
\begin{equation}
\small
    \mathcal{L}_{TF} = \frac{1}{T-k}\sum_{1}^{T-k}\left \| c_t - Concat(\overline{s}_t,\overline{s}_{t+1},...,\overline{s}_{t+k} ) \right \|^{2}_{2}
\end{equation}
where $Concat(\cdot)$ denotes the concatenation of features along the dimensional axe. Unlike the original APC, which operates between the input and output of an autoregressive model, our approach employs a non-streaming ASR model as a teacher to provide semantic information $\mathbf{\overline{s}}$ for guiding this foresight process. This is done to tackle the inherent challenge of obtaining high-quality semantic features from the streaming ASR. After dimensional transformations, the context representation $\mathbf{c}$ is then combined with $\mathbf{h}$ to form the context-enhanced $^c\mathbf{h}$, which is then fed into the acoustic predictor.


Furthermore, since the semantic feature $\{\mathbf{s},\mathbf{\overline{s}}\}$ still may contain speaker-related information. To further ensure the speech decoupling, the bottleneck regulator~\cite{autovcqian2019autovc}, which squeezes out speaker information by reducing dimension size with a linear layer, is applied in $\mathbf{s}$ and $\mathbf{c}$.

\subsection{Training \& Inference Procedure}
\textbf{Training.} During training, the context-enhanced language model and acoustic predictor are trained together. 
The total loss can be described as $\mathcal{L}_{total} = \mathcal{L}_{TF} +\mathcal{L}_{Cont}$ for continuous codec  or $\mathcal{L}_{total} = \mathcal{L}_{TF} + \mathcal{L}_{Disc} $ for discrete version.


\noindent\textbf{Streaming Inference.}
We use the semantic and acoustic features from a short speech clip of the target speaker as the speaker prompt. 
Since this clip is randomly selected, which may contain unfinished pronunciation at the end of the clip, we pad a silence clip after the speaker recording before the conversion process to prevent the unexpected continuation. With this prompt, StreamVoice can stream convert the source speech. In discrete projection, we use greedy decoding to choose the codec token with the highest probability.
Besides, to ensure the real-time streaming inference of StreamVoice, we employ the key-value cache in LM to reduce redundant calculations. 
In practice, since the beginning and end of the source speech can be determined by ASR or voice activity detection (VAD), we don't employ techniques, such as window attention or slide attention, to handle the input. StreamVoice's performance decreases when the long input exceeds the maximum training length. Notably, these techniques can be easily integrated into our framework, providing flexibility for future extensions.

\section{Experiments}


\subsection{Experimental Setup}
\textbf{Corpus. }
A mixed dataset comprising 1,500 hours of Aishell3~\cite{aishell3} and an internal Chinese dataset are used to train StreamVoice and Audiodec~\cite{audiodec}. The internal dataset contains recordings from 2679 Chinese speakers, while we use utterances from 200 speakers in Aishell3. To extract semantic features, we incorporate a streaming ASR Fast-U2++~\cite{fastu2++}, which is implemented by WeNet~\cite{wenet} and trained on WenetSpeech~\cite{wenetspeech}. For zero-shot testing, a set of 400 testing pairs is selected from DIDISpeech~\cite{didispeech} and EMIME~\cite{emime}, each with a source and target speaker utterance. For evaluation of seen speakers, eight speakers from Aishell3 are selected to form 160 conversion pairs. And 3s speech utterance is used as a speaker prompt in inference. The duration of testing utterances is between 3s and 7s.

\noindent\textbf{Implement Details.}
\label{sec:configuration}
We use open-sourced code\footnote{https://github.com/facebookresearch/AudioDec} of Audiodec, which has 4 quantizer layers with a 1024 codebook size and 64 codebook dimension, representing a 24kHz waveform in 20ms frame length. The Fast-U2++ uses an 80ms chunk size to perform streaming inference and compresses a 16kHz waveform into a semantic feature with a 40ms frame length. StreamVoice contains 101M parameters. For context-enhanced LM, we employ the variant of Transformer, LLaMA~\cite{llama}, with 6 layers and 8 heads. The hidden and intermediate sizes are 1024 and 4096. We use the officially released code\footnote{https://github.com/b04901014/MQTTS} to implement the acoustic predictor, which uses a layer Transformer decoder with a hidden size 256, feed-forward hidden size 1024, and 4 heads. In semantic masking, mask ratio $r$ ranges from $0.01$ to $0.02$, and span $l$ is set to 10. The foresight step $k$ is set to 4. The bottleneck regulator compresses feature dimensions by 6 times. During training, the max training length is set to 12s. StreamVoice is trained using 8 V100 GPUs with a batch size of 7 utterances per GPU for 700k steps. We use the AdamW optimizer with a learning rate of $5 \times 10^{-4}$. Exponential decay updates the learning rate after each epoch, using a decay ratio of 0.986.

\noindent\textbf{Evaluation Metrics.}
The mean opinion score (MOS) subjectively measures speech naturalness (NMOS) and speaker similarity (SMOS), which are calculated with 95$\%$ confidence intervals. We randomly select 120 testing pairs for subjective evaluations involving a group of 15 listeners. For objective evaluations, a neural network-based system with open-source implementation\footnote{https://github.com/AndreevP/wvmos} is used to measure speech quality (WV-MOS). Character error rate (CER) measured by an ASR model\footnote{https://github.com/wenet-e2e/wenet/tree/main/examples/wenetspeech\label{model:nsasr-wenetspeech}} indicates the speech intelligibility. Speaker similarity (SSIM) is calculated by an SV model~\cite{ECAPA_TDNN} to determine if the converted speech matches the target speaker. Real-time factor (RTF) and latency indicate the streaming performance. 


\begin{table}[htp]
\footnotesize
\setlength{\tabcolsep}{0.2mm}
\renewcommand\arraystretch{1.22}
\begin{tabular}{lcccccc}
\hline
\multicolumn{1}{l}{\multirow{2}{*}{Method}} & \multicolumn{3}{c}{Quality} & \multicolumn{2}{c}{Similarity}  \\ \cline{2-6}
  &  \scriptsize{NMOS} $\uparrow$  & \scriptsize{WVMOS} $\uparrow$  & \scriptsize{CER}  $\downarrow$ & \scriptsize{SMOS} $\uparrow$ & \scriptsize{SSIM} $\uparrow$ \\ \hline
 GT~(origin) & -  & 3.61  & 6.29  & - &  0.803 \\ \hline
 \multicolumn{2}{l}{\textit{\textcolor[RGB]{57,57,57}{Non-streaming Topline}}}    &   &   &  &   \\ 
 LM-VC & 3.80$\pm$0.09  & 3.74  &  8.93 & 3.78$\pm$0.08   & 0.742   \\
 NS-StreamVoice & 3.87$\pm$0.07  & 3.68  & 8.51 & 3.73$\pm$0.11  & 0.755   \\ \hline
 \multicolumn{2}{l}{\textit{\textcolor[RGB]{57,57,57}{Streaming Model}}}    &   &   &  &   \\ 
 C-StreamVoice & 3.72$\pm$0.10  & 3.49  & 10.2 & 3.67$\pm$0.09  & 0.729  \\
  \makecell[l]{StreamVoice} & 3.83$\pm$0.09 & 3.63  &  9.43    & 3.74$\pm$0.08  & 0.740  \\ \hline
\end{tabular}
\caption{Zero-shot performance (unseen speakers)}
\label{exp:zeroshot}
\end{table}

\subsection{Experiments Results}

\subsubsection{Zero-shot Evaluation}
To evaluate the zero-shot VC performance, one recent LM-based zero-shot VC system, \textit{LM-VC}~\cite{LM-VC}, is selected as the topline system. Besides, a variant of StreamVoice, referred to as \textit{NS-StreamVoice}, using a non-streaming ASR for semantic extraction, is also compared. We implement the proposed system \textit{StreamVoice} integration discrete projection, while \textit{C-StreamVoice} also involves the evaluation since speech can represented in continuous form by codec model. Table.~\ref{exp:zeroshot} presents both subjective and objective results. Compared with the non-streaming topline LM-VC, our proposed StreamVoice can achieve close results regarding subjective NMOS and SMOS, while a performance gap still exists. Similar results are also observed in objective results. The non-streaming StreamVoice even surpasses the topline model in certain aspects, indicating the effectiveness of our streamable architecture for zero-shot VC. Additionally, C-StreamVoice exhibits inferior performance compared to the discrete version, which can contribute to the over-smoothness in speech generation~\cite{ren2022revisiting} and the mismatch between the ground truth and predicted features. 

As illustrated in Table.~\ref{exp:speed}, the RTF of the entire pipeline is below 1, which meets the real-time requirement. Consisting of chunk-waiting latency (80ms) and model inference latency, the overall pipeline latency is 124.3 ms. If using a V100 GPU, StreamVoice can obtain an RTF of 0.56, and the overall latency reaches 137.2 ms. Importantly, unlike previous streaming VC, our VC model is entirely causal without any future look-ahead, highlighting its powerful modeling capability. These results show that StreamVoice can achieve high-quality zero-shot VC in streaming scenarios.


\begin{table}[htp]
\footnotesize
\setlength{\tabcolsep}{1.5mm}
\renewcommand\arraystretch{1.3}
\centering
\begin{tabular}{lcl}
\hline
   & RTF  & Latency~(ms)   \\ \hline
  ASR Encoder  & 0.13     & \quad10.4  \\ 
   Codec Decoder   & 0.004     & \quad0.3   \\
  StreamVoice   & 0.42 & \quad33.6  \\ \hline
 Overall   & 0.554 &  \quad44.3+80=124.3   \\ \hline
\end{tabular}
\caption{Speed tested on an A100 80G GPU. Latency is obtained by multiplying RTF
by 80ms chunk size.}
\label{exp:speed}
\end{table}

\begin{table}[htp]
\footnotesize
\setlength{\tabcolsep}{0.3mm}
\renewcommand\arraystretch{1.1}
\begin{tabular}{lccccc}
\hline
\multicolumn{1}{l}{\multirow{2}{*}{Method}} & \multicolumn{3}{c}{Quality} & \multicolumn{2}{c}{Similarity}  \\ \cline{2-6} 
  &   \scriptsize{NMOS} $\uparrow$  & \scriptsize{WVMOS} $\uparrow$  & \scriptsize{CER}  $\downarrow$ & \scriptsize{SMOS} $\uparrow$ & \scriptsize{SSIM} $\uparrow$ \\ \hline
 GT~(origin)   & -   & 3.65  & 6.29  &  - & 0.729  \\ \hline
 \multicolumn{3}{l}{\textit{\textcolor[RGB]{57,57,57}{Non-streaming Topline}}}       &   &  &   \\ 
 NS-VC  &  3.85$\pm$0.09 & 3.71  & 8.39 & 3.92$\pm$0.08   & 0.744   \\
 \hline
 \multicolumn{3}{l}{\textit{\textcolor[RGB]{57,57,57}{Streaming Model}}}       &   &  &   \\ 
 IBF-VC  & 3.71$\pm$0.09  &  3.48  &  9.52  & 3.67$\pm$0.10   & 0.687   \\
 DualVC2   & 3.80$\pm$0.10  & 3.57  &  10.2 &  3.85$\pm$0.09  & 0.703  \\
  StreamVoice  & 3.82$\pm$0.09 & 3.50  & 10.0  & 3.82$\pm$0.10  & 0.694 \\
  $\quad  +$  Tuning  & 3.78$\pm$0.08 & 3.52  & 10.4    &  3.87$\pm$0.10 & 0.714  \\
 \hline
\end{tabular}
\caption{In-dataset performance (seen speakers)}
\label{exp:seenspk}
\end{table}

\subsubsection{In-dataset Evaluation}
To get further insight into StreamVoice, we conducted an in-dataset evaluation on eight seen speakers, as shown in Table.~\ref{exp:seenspk}. A non-streaming VC system~\cite{nusnwpu} achieving any2many VC, is selected, referred to as \textit{NS-VC}. Also, \textit{IBF-VC}~\cite{chen2022streaming} and \textit{DualVC2}~\cite{dualvc2} are recently proposed streaming models for any2many VC. As observed, a performance gap exists between the strong non-streaming topline and streaming models. Among the streaming models, StreamVoice, designed for the zero-shot scenario, delivers similar results to systems designed for in-dataset speakers, even though StreamVoice uses a smaller chunk size of 80ms in streaming ASR, achieving lower ASR performance. In contrast, IBF-VC and DualVC2 employ 160ms chunk size of ASR for streaming VC. It indicates StreamVoice's good conversion ability. With available utterances of target speakers, fine-tuning yields superior performance. This indicates our system can be easily applied to various scenarios with or without the utterances of target speakers.



\begin{table}[htp]
\footnotesize
\setlength{\tabcolsep}{0.6mm}
\renewcommand\arraystretch{1.3}
\centering
\begin{tabular}{lccc}
\hline
 Method   & WVMOS $\uparrow$  & CER  $\downarrow$  & SSIM $\uparrow$ \\ \hline
  StreamVoice   & 3.63     & 9.43  & 0.740 \\ \hline
 \multicolumn{3}{l}{ $\quad$ \textit{\textcolor[RGB]{57,57,57}{w/o Teacher-guided Context Foresight}}}           &   \\ 
 $\quad$ $\quad$ w/o $\mathcal{L}_{TF}(\overline{s}_t)$    & 2.56  & 76.8   & 0.59   \\
 $\quad$ $\quad$ w/o $\mathcal{L}_{TF}(\overline{s}_{t+1:t+k})$   & 3.39 & 13.7  & 0.728   \\ \hline
 $\quad$w/o Semantic Masking   & 3.47 & 13.0  & 0.715  \\\hline
 $\quad$w/o Bottleneck Regulator   & 3.59  & 9.21  & 0.718 \\\hline
\end{tabular}
\caption{Results of ablation studies.}
\label{exp:abs}
\end{table}

\subsection{Ablation Study}

As presented in Table~\ref{exp:abs}, we conducted several ablations studies.
In \textit{w/o} teacher-guided context foresight, we discard the prediction of current and future semantic information, forming two ablations \textit{w/o} $\mathcal{L}_{TF}(\overline{s}_t)$ and \textit{w/o} $\mathcal{L}_{TF}(\overline{s}_{t+1:t+k})$. As can be seen, a noticeable decrease occurs in all evaluation metrics when the $\mathcal{L}_{TF}(\overline{s}_{t+1:t+k})$ is discarded, especially in WVMOS and CER. This indicates that this foresight improves performance in capturing the linguistic content. But when only integrating context from future semantics, the model \textit{w/o} $\mathcal{L}_{TF}(\overline{s}_t)$ faces severe performance loss. It shows that only using future information interferes with delivering current linguistic content.
In \textit{w/o} semantic masking, we observe a performance decrease in all evaluation metrics when the semantic masking is discarded. This indicates that StreamVoice, trained with semantic masking, effectively enhances contextual learning from the preceding input while improving speaker timbre capturing.
Furthermore, the results of dropping the bottleneck regulator show that its integration effectively prevents the source speaker information contained in the semantic feature from leaking into the converted speech, with little effect on speech quality.





\begin{table}[htp]
\footnotesize
\setlength{\tabcolsep}{0.6mm}
\renewcommand\arraystretch{1.3}
\centering
\begin{tabular}{lccc}
\hline
 Type of ASR   & WVMOS $\uparrow$  & CER  $\downarrow$  & SSIM $\uparrow$ \\ \hline
  Non-streaming ASR   & 3.68     & 8.51  & 0.755 \\ \hline
 \multicolumn{4}{l}{\textit{\textcolor[RGB]{57,57,57}{Streaming ASR ~\cite{streamctc}}}}          \\ 
 $\quad +$ 0ms Future Look-ahead    & 3.19 &  91.7 & 0.674   \\
 $\quad +$ 160ms Future Look-ahead   & 3.48 &  10.6    & 0.727   \\ \hline
  \multicolumn{4}{l}{\textit{\textcolor[RGB]{57,57,57}{Streaming ASR (Fast-U2++~\cite{fastu2++})}}}            \\ 
 $\quad$ Chunk~(80ms)   & 3.63     & 9.43  & 0.740 \\
 $\quad$ Chunk~(160ms)   & 3.69     & 9.16  & 0.744 \\\hline
\end{tabular}
\caption{Analysis of dependency on different ASR.}
\label{exp:asr}
\end{table}


\subsection{Discussion: Dependency Analysis}
\label{sec:dependency}


In this section, we will explore the dependency relations between the selection of ASR and codec and the performance of StreamVoice.

\textbf{ASR.} 
To investigate the impact of ASR on StreamVoice, three representative ASR systems, including non-streaming ASR~\footref{model:nsasr-wenetspeech}, widely used CTC-based streaming ASR~\cite{streamctc}, and the recently proposed streaming Fast-U2++~\cite{fastu2++}, are selected to perform semantic extraction. As can be seen in Table~\ref{exp:asr}, StreamVoice using semantic features of non-streaming ASR outperforms those using streaming ASR. This discrepancy may be attributed to the inherent performance gap between non-streaming and streaming ASR models, resulting in different semantic extraction abilities. Besides, without future look-ahead in StreamVoice, using semantic features from \cite{streamctc} cannot achieve reasonable conversion, while we introduce a 160ms future look-ahead in StreamVoice, i.e., modeling $p(a_t|\mathbf{a}_{1:t-1},\mathbf{s}_{1:t+m},t)$ with $m$ future look-ahead, yield good conversion results. This issue may arise from delayed CTC spike distributions and token emission latency existing in streaming ASR~\cite{fastu2++}, leading to semantic information shifting. Benefiting from the low emission latency of Fast-U2++, StreamVoice can perform conversion without future look-ahead. With a longer chunk size employed in Fast-U2++, StreamVoice can obtain better results while reaching a larger latency of 270ms. A trade-off still exists between performance and speed.

\begin{table}[htp]
\footnotesize
\setlength{\tabcolsep}{0.6mm}
\renewcommand\arraystretch{1.3}
\centering
\begin{tabular}{lcccc}
\hline
  Type of Audiodec  & WVMOS $\uparrow$  & CER  $\downarrow$  & SSIM $\uparrow$  &RTF\\ \hline
 w/ $2kbps$ Audiodec    & 3.63 & 9.43  & 0.740   & 0.42 \\
 w/ $8kbps$ Audiodec   & 3.61 &  9.38  & 0.738  & 0.61 \\ 
 Large w/ $8kbps$ Audiodec     & 3.68 &  9.12 & 0.751 & 0.90  \\ \hline
\end{tabular}
\caption{Analysis of dependency on Audiodec with various bitrate.}
\vspace{-10pt}
\label{exp:codec}
\end{table}

\textbf{Codec.}
In StreamVoice, we employ a low-latency streaming codec Audiodec~\cite{audiodec}. As presented in Table.~\ref{exp:codec}, we validate the performance of StreamVoice using codecs with different bitrates, including 2kbps and 8kbps, where higher bitrate codecs achieve superior reconstruction quality to lower bitrate ones. The 2kbps Audiodec utilizes 4 layers of quantization and represents audio with a frame length of 20ms, while the 8kbps Audiodec employs 8 layers with a frame length of 10ms. Using the configuration of StreamVoice mentioned in Section \ref{sec:configuration}, the results in different bitrates of codec models show no obvious differences. When increasing the number of transformer layers in the codec predictor, forming \textit{Large w/ 8kbps Audiodec}, conversion performance using 8kbps codec improves noticeably, but resulting in slower inference. This result shows that the design of StreamVoice depends on the codec configuration, affecting both conversion quality and inference speed.



\section{Conclusions}

This paper introduces StreamVoice, a novel LM-based zero-shot VC system designed for streaming scenarios.
Specifically, StreamVoice employs a single-stage framework encompassing a context-aware LM and an acoustic predictor. The casual design of the model's input and structure ensures compliance with streaming behavior. To address performance degradation caused by missing complete contextual information in streaming scenarios, context-aware LM adopts teacher-guided context foresight to make the model able to forecast the current and future information given by a teacher. Besides, semantic masking is introduced in LM to enhance context learning from historical input and facilitate better disentanglement. Finally, an acoustic predictor collaborates with the LM to generate the target speech. Experiments demonstrate that StreamVoice achieves streaming zero-shot VC while maintaining performance comparable to non-streaming VC systems. 

\section{Limitations}  
We have to point out that StreamVoice still has limitations. In our configuration, StreamVoice needs a GPU, such as V100 and A100, to achieve real-time streaming inference. The design of streaming VC heavily relies on the ASR and the speech codec as mentioned in Section~\ref{sec:dependency}. Besides, StreamVoice also faces the out-of-domain problem, which causes performance degradation for utterances with accents, strong emotions, or unseen recording environments. Our future work will first use more training data with diversity coverage to explore StreamVoice's modeling ability. Also, we will focus on optimizing our streaming pipeline, such as high-fidelity codec with low bitrate and unified streaming model.

\section{Ethics Statement}
Since StreamVoice can convert source speech to desired speakers, it may carry potential risks of misuse for various purposes, such as spreading fake information or phone fraud. To prevent the abuse of the VC technology, many studies have focused on synthetic speech detection~\cite{ADD}. Meanwhile, we also encourage the public to report the illegal usage of VC to the appropriate authorities.

\bibliography{custom}

\appendix



\end{document}